\documentclass[num,preprint,12pt]{elsarticle}

\usepackage{amssymb}
\usepackage{amsthm}
\usepackage{amsmath}
\usepackage{array}
\usepackage{dsfont}
\usepackage{wasysym}
\usepackage{dutchcal}
\usepackage[squaren]{SIunits}
\usepackage{subfig}
\usepackage[export]{adjustbox}
\usepackage{rotating}
\usepackage[dvipsnames]{xcolor}
\usepackage{hyperref}

\definecolor{AS}{rgb}{1,0,1}

\usepackage{mathtools}
\usepackage{lineno}
\usepackage{ulem}

\newcommand{\Capi}{\operatorname{\mathrm{C\kern-.08em a}}} 
\newcommand{\Cael}{\operatorname{\mathrm{C\kern-.08em a_{El}}}} 
\newcommand{\Gp}{G^\prime} 
\newcommand{\Gpp}{G^{\prime\prime}} 
\newcommand{\dd}{\mathrm{d}}
\newcommand{\ii}{\mathrm{i}}

\journal{Soft Matter}

\begin{document}

\begin{frontmatter}
\title{Rheological response of soft Solid/Liquid Composites}

\author[1,2]{Elina Gilbert}

\author[3]{Anniina Salonen}

\author[1]{Christophe Poulard\corref{cor1}}
\ead{christophe.poulard@universite-paris-saclay.fr}

\affiliation[1]{organization={UMR8502 Laboratoire de Physique des Solides, Université Paris-Saclay},
    addressline={1 rue Nicolas Appert, bâtiment 510},
    postcode={91400},
    city={Orsay},
    country={France}}
    
\affiliation[2]{organization={SPEC – Service de Physique de l'État Condensé,
        CEA, CNRS, Université Paris-Saclay},
    addressline={Bâtiment 772, Orme des Merisiers},
    postcode={91191},
    city={Gif-sur-Yvette},
    country={France}}
    
\affiliation[3]{organization={Soft Matter Sciences and Engineering, ESPCI Paris, PSL University, CNRS, Sorbonne Université},
    addressline={10 rue Vauquelin},
    postcode={75005},
    city={Paris},
    country={France}}

\cortext[cor1]{Corresponding author}


\begin{abstract}
Understanding a material's dissipative response is important for their use in many applications, such as adhesion or fracture resistance. In dispersions, the interplay between matrix and inclusions complicates any description. Fractional rheology is conveniently used to fit the storage and loss moduli of complex materials. In conjugation with superposition methods, they allow to better capture the behavior of materials of complex rheology.
We study the rheology of soft solid/liquid composites of liquid poly(ethylene glycol) (PEG) droplets in a soft poly(dimethylsiloxane) (PDMS)  matrix. We analyze the influence of the droplets through fractional rheology and a time-concentration superposition in the continuous-phase-dominated region. Viscous dissipation increases proportionally with volume fraction, independently of the frequency, whereas the elastic response is almost unchanged.
\end{abstract}

\end{frontmatter}

\section{Introduction}
Before even the foundation of the Society of Rheology in 1929, one of the main concerns of soon-to-be rheologists was to describe the flow behavior of complex materials using the simplest possible models, and to link those models to physical-chemical parameters of the studied systems. In 1921, Nutting proposed a logarithmic model to improve previous descriptions by Michelson on viscoelastic materials (then called "elastico-viscous") \cite{nutting_new_1921,michelson_laws_1917}, leading to a power-law deformation curve:
\begin{equation*}
	\frac{\dd\gamma}{\gamma} \propto n\frac{\dd t}{t} + m\frac{\dd \tau}{\tau} 
\end{equation*}
where $\gamma$ is the shear strain, $t$ the experimental time and $\tau$ the shear stress. The experimental observations behind this empirical law also led Nutting to suggest a relationship between solicitation time and temperature of the material, thirty years before the well known Williams-Landel-Ferry equation \cite{williams_temperature_1955}. About twenty years later, Blair and his collaborators came back to these equations while trying to describe the firmness and springiness of cheese, as the elastic modulus and the viscosity proved insufficient \cite{blair_subjective_1939}. Blair proposed a "principle of intermediacy" for complex materials with behaviors in-between Newtonian liquids and Hookean solids, with a non-unit derivative, such that:
\begin{equation*}
	\tau(t) = \mathds{V}\frac{\dd^\alpha\gamma}{\dd t^\alpha}
	\label{eq:fractional}
\end{equation*}
where $\alpha\in[0;1]$ and $\mathds{V}$ in $\pascal.\second^{\alpha}$ is a "quasi-property" of the material \cite{blair_analytical_1944,blair_limitations_1947}. When $\alpha=0$, the material is fully elastic and $\mathds{V}$ is the shear modulus $G$. When $\alpha=1$, it is a viscous liquid, and $\mathds{V}$ is the viscosity $\eta$. Between those extreme behaviors, $\alpha$ represents a relative importance of the dissipative character of the material, and $\mathds{V}$ characterizes an integrated view of dynamic relaxations in the sample relating to its structure.

It took another twenty years until Caputo formally defined the fractional derivative \cite{caputo_linear_1967}:
\begin{equation*}
	\frac{\dd^\alpha f}{\dd t^\alpha} = \frac{1}{\Gamma\left(1-\alpha\right)}\int_0^t\frac{f(x)}{(t-x^\alpha)}\dd x
	\label{eq:caputo}
\end{equation*}
with the gamma function $\Gamma(x) = \int_0^\infty t^{x-1}e^{-z}\dd z$.

The introduction of the Fourier transform of the fractional model yet another twenty years later, allowed for its use in oscillatory models.
\begin{equation*}
	\tau^*(\omega) = \mathds{V}(\ii\omega)^\alpha
	\label{eq:oscfrac}
\end{equation*}
This allowed for physical analysis of the fractional model which has been convenient to compare engineered materials and to implement in finite element models, but with little physical basis. Bagley and Torvik established an analogy between a fractional model and a Rouse polymer represented by an extended Maxwell model: integrating over all the branches of the model led to a frequency dependency to the power $1/2$ in the complex modulus \cite{bagley_theoretical_1983,bagley_fractional_1986}. Gloeckle and Nonnenmacher used a fractional Zener model with a fractional integration method to fit experimental data on polymer melts \cite{nonnenmacher_fractional_1991}. This model showed fractal time processes in the relaxation spectrum \cite{gloeckle_fractional_1991,gloeckle_fractional_1994} whose study inspired another model, similar to an extended Maxwell model but with a recursive definition of each branch in a ladder structure \cite{schiessel_hierarchical_1993,schiessel_generalized_1995}. 

Due to the complexity of the processes at play no unified theory linking fractional models to the structure of complex systems has yet emerged, however a recent series of papers use the fractal ladder model and its fractional approximation to study the behavior of colloidal gels. The model shows a link between the fractal dimension in the structural heterogeneities and the rheological behavior of the theoretical gels \cite{jaishankar_fractional_2014,jaishankar_power_2013,bantawa_hidden_2023}, and links nicely with Blair's original study on the texture of cheese \cite{faber_describing_2017, faber_describing_2017a, faber_firm_2017}.

In the last decades, solid-liquid composites (SLC), particularly those consisting of elastomer matrices with liquid droplet inclusions, have emerged as a promising class of materials that combine the structural integrity of solids with the properties of liquids \cite{style_solid–liquid_2020}. These innovative systems offer opportunities for developing multi-functional materials with tailored characteristics, making them increasingly important in various fields of materials science and engineering \cite{sheng_solid–liquid_2021,yu_solid–liquid_2021,li_solid–liquid_2023, cai_printing_2020}. The mechanical behavior of such systems leads to interesting properties in fields such as food science of course \cite{langendorfer_viscoelastic_2024}, but also soft actuators \cite{ford_multifunctional_2019} and soft electronics \cite{ma_soft_2017, yun_liquid_2019,ankit_high_2020,peng_highly_2023} using liquid metal or ionic liquids in the inclusions. Many authors have focused on describing the stiffness of such materials, taking into account the interplay between the capillary pressure in the droplets and the elasticity of the solid matrix. Special care was taken because of the uncertainties associated to the rheological measurements on such complex systems \cite{style_stiffening_2014,ghosh_elastomers_2022,ghasemi_multiscale_2021,hamdia_quantifying_2022}, but little attention has been given to their dissipative behavior.
In the present study, we focus on soft SLC, composed of encapsulated liquid droplets in a soft solid continuous phase, which we also call "solid emulsions".
We fit these SLC using fractional rheology, and use superposition methods to link the model to the response of the system, focusing on the influence of the liquid droplets on viscous dissipation.

\section{Materials and methods}
\subsection{Generation of the solid emulsions}
The continuous phase is a mix of two commercially available PDMS elastomer kits, Sylgard 184 and Sylgard 527, provided by Dow\textsuperscript{\tiny\textregistered}. Both kits are prepared according to specification, then they are mixed together at a proportion $80:20$ Sylgard527:Sylgard184. A block copolymer of PDMS and PEG, DBE-224, provided by Gelest\textsuperscript{\tiny\textregistered}, is added to the liquid continuous phase to stabilize the emulsion. Pure liquid PEG 600 (Sigma-Aldrich\textsuperscript{\tiny\textregistered}) is dispersed in the continuous phase mix using an IKA T-18 Basic Ultra-Turrax\textsuperscript{\tiny\textregistered} with an SN18-10G probe.
The dispersion step generates heat which catalyzes the cross-linking of the PDMS. However, PEG 600 crystallizes just below room temperature and cooling the emulsion too intensely could break it. Therefore, the emulsion is generated in a sequential manner. During the first two-minutes-long mixing step, on speed 3\footnote{The T-18 Basic Ultra Turrax only has nominal speeds, which correspond to a given power of the motor, and no control over the rotation-per-minute which depends on the viscosity of the mixed liquids.}, the emulsion is cooled in an ice bath for 10 seconds every 30 seconds. After homogenization by hand, the second dispersion step is performed on speed 1 for one minute, with again two cooling steps in an ice bath, of 10 seconds each.
The homogeneous emulsions are then poured in a Petri dish mold ($\diameter35~\milli\metre$ or $\diameter25~\milli\metre$) and left to cure at room temperature for two to three days, allowing for the evacuation of most of the unwanted air bubbles created by the dispersion step.
The resulting emulsions, which were characterized in detail in a previous study, are highly polydisperse. However, the droplet size distribution is very similar between samples, with droplet sizes ranging from about $40~\nano\metre$ to $20~\micro\metre$, and a distribution peak around $1~\micro\metre$ \cite{gilbert_decoupling_2024}.

\subsection{Linear oscillatory rheometry}
\begin{figure}[hbt!]
    \centering
    \includegraphics[width=7cm]{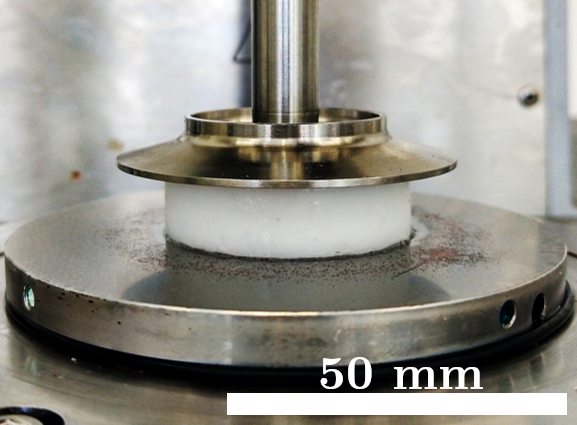}
    \caption{\label{fig:rheo}
        Photograph of a solid emulsion on the rheometer in plate-plate configuration.}
\end{figure}

Oscillatory rheology measurements are conducted on Anton Paar\textsuperscript{\tiny\textregistered} (MCR301 and MCR302) rheometers, in plate-plate configuration, on a Peltier surface, as shown on figure \ref{fig:rheo}.

The samples have a diameter of either $25~\milli\metre$ or $35~\milli\metre$ and the measuring tool geometry is chosen to be as close as possible in diameter. For the larger samples, a P50 was chosen, and the measured moduli were corrected by a factor of $\left(R_\text{measuring tool}/R_\text{sample}\right)^4$ to account for the difference in diameter. This is the case shown in Figure \ref{fig:rheo}. For the smaller samples, a P25 was chosen and no correction was needed. To ensure contact during the measurements, the initial normal force is set at $2~\newton$ and a thin layer of the continuous phase mix is crosslinked between the sample and the two plates of the rheometer at $70\degreecelsius$ for $4~\hour$. The frequency sweeps in this study are then performed at $25\degreecelsius$ at fixed deformations of $\gamma=0.1\%$ or $\gamma=0.01\%$.

\section{Results and Discussion}
\subsection{Frequency-dependent behavior of liquid-in-solid emulsions}
\begin{figure*}[hbt!]
	\hspace{-1cm}
    \includegraphics[width=15.75cm,keepaspectratio]{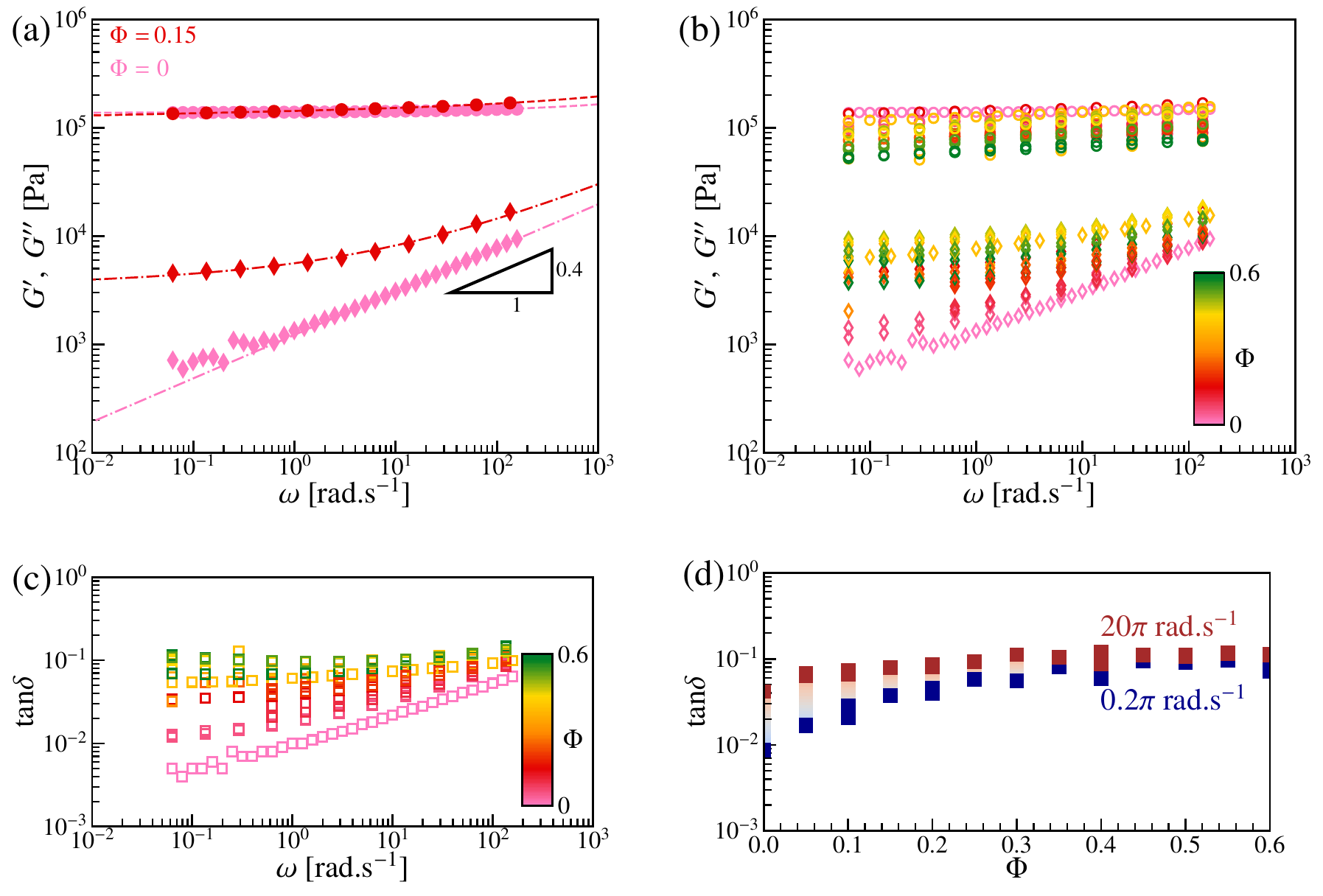}
	\caption{\textbf{(a)} Comparison of the frequency dependency of the storage ($\CIRCLE$) and loss ($\blacklozenge$) moduli of the pure continuous phase (pink, $\Phi=0$), and a solid emulsion of volume fraction $\Phi=0.15$ (red). %
        \textbf{(b)} Evolution of the storage ($\Circle$) and loss ($\lozenge$) moduli of solid emulsions with the angular frequency. The colors correspond to the volume fraction of each sample.%
        \textbf{(c)} Evolution of the loss factor of solid emulsions with the angular frequency. %
        \textbf{(d)} Evolution of the loss factor with the volume fraction of solid emulsions at two angular frequencies, $20\pi~\radian.\second^{-1}$ (red) and $0.2\pi~\radian.\second^{-1}$ (blue). The lighter colors represent the range of behaviors between these frequencies.
		\label{fig:brut}}
\end{figure*}

We make SLC with various volume fractions of liquid polyethylene glycol (PEG) inclusions embedded in a viscoelastic poly dimethylsiloxane (PDMS) matrix. The resulting rheological response of the samples is shown in Figure \ref{fig:brut}.

In small deformation oscillatory rheometry, all the liquid-in-solid emulsions show a typical rheological response similar to the one shown in Figure \ref{fig:brut}(a). The storage modulus $\Gp$ is close to the elastomeric plateau of the PDMS continuous phase (in pink), but shows a weak increase. The loss modulus $\Gpp$ shows two asymptotic behaviors at low and high frequencies. At high frequency, the asymptotic behavior tends towards a power law with a non-integer exponent of $0.4$ of the pure elastomer continuous phase. At low frequency, the influence of the inclusion of droplets is visible, flattening the loss modulus towards a power-law of very small exponent. 
The weak increase in the storage modulus seems to follow a similar power-law of small exponent. These asymptotic behaviors are similar at all volume fractions as shown on Figure \ref{fig:brut}(b). They translate into a transition in the loss factor $\mathrm{tan}\delta$  between a low plateau at low frequencies and an increase at higher frequencies, as shown on Figure \ref{fig:brut}(c).

While the evolution in storage modulus shows no specific trend as we previously observed \cite{gilbert_decoupling_2024}, the plateau value of the loss modulus increases with volume fraction for $\Phi\le0.5$. The frequency of transition between the two asymptotic behaviors in the loss modulus is changing with volume fraction. This translates into different frequencies of transition in the loss factor. The behavior at high frequency is very similar across all volume fractions, as Figure \ref{fig:brut}(d) emphasizes, which can be interpreted as the continuous phase relaxation completely overtaking the relaxation of the droplets. Indeed, at high frequencies, the elastic continuous phase will "respond" more quickly to solicitations than the liquid droplets.

\begin{figure*}[hbt!]
    \hspace{-1cm}
    \includegraphics[width=15.75cm]{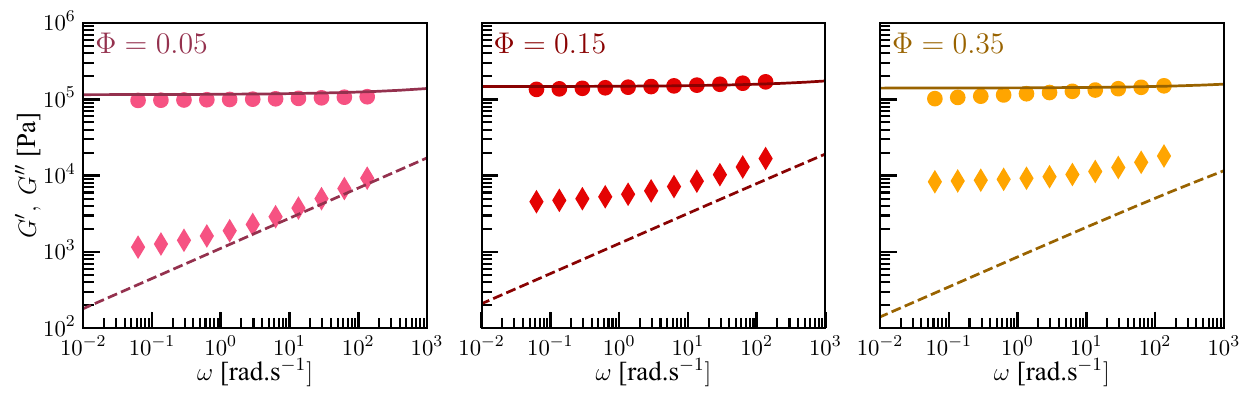}
    \caption{Fit by Palierne's model \cite{palierne_linear_1990} of the storage ($\CIRCLE$) and loss ($\blacklozenge$) moduli of three solid emulsions of volume fraction $\Phi=0.05$, $0.15$ and $0.35$.}
    \label{fig:Paltest}
\end{figure*}

A first idea to have a better grasp of the interplay of the continuous and dispersed phase over the moduli would be to use Palierne's seminal equation for blends of viscoelastic materials \cite{palierne_linear_1990}. It has however been shown that for these specific solid emulsions, Palierne's model failed to capture in a satisfactory manner both the elastic and the dissipative response \cite{gilbert_decoupling_2024}. Even the application of a volume fraction correction to allow the model to encompass higher volume fractions \cite{pal_new_2008}, does not allow proper fitting in frequency as represented on figure \ref{fig:Paltest}.

While the storage modulus is relatively well-captured by Palierne's model, the flattening of the loss modulus at low frequencies is not fitted at all. This would nevertheless be an interesting result if the continuous-phase-dominated region was well-described by the model, however, while the high frequency behavior of the loss modulus for the sample at small volume fraction $\Phi=0.05$ is still correctly fitted by the model, for higher volume fractions a larger discrepancy appears. Indeed, in the relevant orders of magnitude of droplet radius and interfacial tension (respectively $10^{-7}\--10^{-5}~\metre$ and $10^{-3}\--10^{-2}~\newton.\metre^{-1}$ \cite{gilbert_decoupling_2024}), Palierne's model tends to predict that the loss modulus decreases with volume fraction. In our experiments however, the loss modulus increases with volume fraction. Thus, we find another solution to study the full rheological responses of solid emulsions.

Given the two clear non-integer exponents in the asymptotic power-laws in the dissipative response of our samples, we choose to model the solid emulsions using fractional rheology. The aim is to represent our data with the simplest possible model, so we use a modified Kelvin-Voigt model with fractional branches to represent the two asymptotes.

\subsection{Fractional Kelvin-Voigt model}
\begin{figure*}[hbt!]
    \centering
    \subfloat{(a)}\subfloat{
        \includegraphics[width=8cm,keepaspectratio,valign=t]{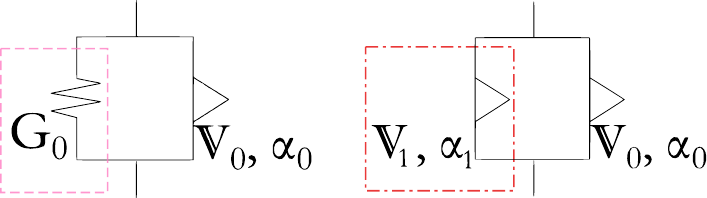}}\\
    \subfloat{\hspace{-1em}(b)\hspace{-1em}}\subfloat{
        \includegraphics[width=14cm,keepaspectratio,valign=t]{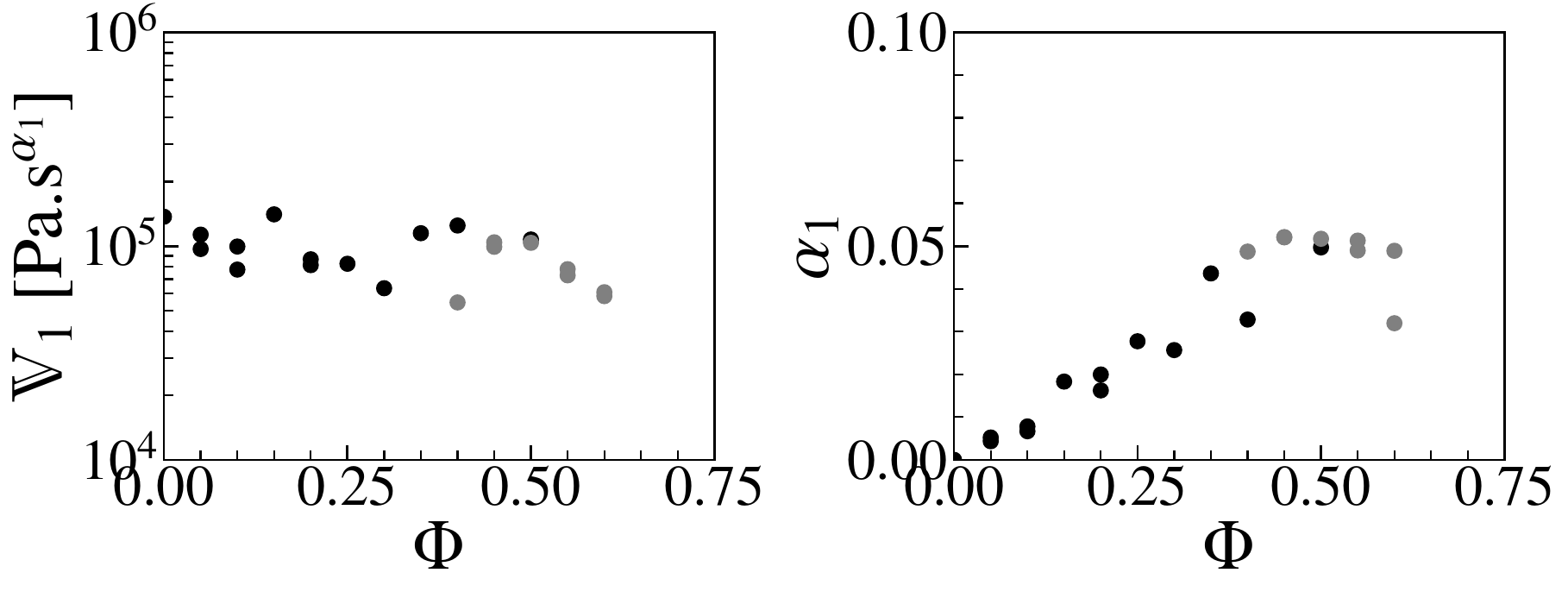}}\\
    \caption{\textbf{(a)} Fractional rheology models used to represent (\textbf{Left}) the PDMS (FKV0) and (\textbf{Right}) the solid emulsions (FKV1). The changing branch corresponds to the different power-laws visible at low frequencies and on $\Gp$. $\mathds{V}_0$ is set at $2.06~\pascal.\second^{\alpha_0}$ and $\alpha_0$ at $0.4$. %
        \textbf{(b) Left }Evolution of the quasi-property $\mathds{V}_1$ with the volume fraction. The grey circles correspond to samples for which $R^2<0.9$. %
            \textbf{Right} Evolution of the fractional power-law exponent $\alpha_1$ with  volume fraction.
        \label{fig:FKV}}
\end{figure*}

Elastomers can usually be represented by a fractional Kelvin-Voigt model in the frequency interval of our study. Their fractional properties come from the presence of free and dangling chains in the bulk leading to dynamic relaxations \cite{curro_theoretical_1983}. The storage modulus is on the elastomeric plateau while the loss modulus is increasing following a power-law with a non-integer exponent \cite{sollich_rheology_1997,sollich_rheological_1998,hebraud_role_2000}. Thus, we can already fit the silicone continuous phase with a fractional Kelvin-Voigt model, with one purely elastic branch for the storage modulus, and a fractional branch for the loss modulus, which we will name FKV0 for the continuation of this paper. This gives us a first model of the form
\begin{equation}
	G^*_0 = \mathds{V}_0(\ii\omega)^{\alpha_0} + G_0
	\label{eq:FKV0}
\end{equation}
We extend this model to our solid emulsion samples, by changing the branch at play for the low-frequency behavior into a second fractional branch, in order to have two power-law behaviors (named FKV1), as follows:
\begin{equation}
	G^* = \mathds{V}_0(\ii\omega)^{\alpha_0} + \mathds{V}_1(\ii\omega)^{\alpha_1}
	\label{eq:FKV1}
\end{equation}
The schematic representation of these models as well as the resulting fitting parameters are shown on Figure \ref{fig:FKV}.

As shown on Figure \ref{fig:FKV}(a), the fractional branch for the PDMS ($\mathds{V}_0$, $\alpha_0$) is retained for all the solid emulsion samples. To fit them with FKV1, we set the values of $\mathds{V}_0$ and $\alpha_0$ at respectively $2.06~\pascal.\second^{\alpha_0}$ and $0.4$. The quasi-property $\mathds{V}_1$ represented on Figure \ref{fig:FKV}(b), corresponding to the plateau value in loss modulus at low frequencies, seems to stay constant with volume fraction for $\Phi\le0.5$. At higher volume fraction, despite a lower quality of fit, we can observe a decrease of $\mathds{V}_1$, similarly to the behavior of the storage moduli measured at $\omega=2\pi~\radian.\second^{-1}$, showing that the quasi-property is strongly coupled to the elastic response of the samples. The power-law exponent $\alpha_1$, however, increases linearly with the volume fraction for $\Phi\le0.5$, while keeping a very low value below $0.1$.

We can interpret $\alpha_1$ as a measure of the importance of viscous dissipation compared to a purely elastic response. The higher its value, the more prominent a role viscosity plays. In this case, it seems that the inclusion of droplets mainly impacts system elasticity, however their slight dissipative behavior is extremely important to describe the full rheological behavior of the solid emulsions. Indeed, if the inclusions were only elastic, there would be no impact on the loss modulus of the emulsions, instead of the asymptote observed at low frequencies. The variation in loss moduli are interesting, as bulk dissipation in elastomeric materials drives their adhesive behavior and their fracture resistance \cite{creton_fracture_2016}, making solid emulsions especially promising soft adhesives or impact resistance materials \cite{yu_solid–liquid_2021}. In fractional materials, the exponent characterizes microscopic dynamic relaxation events. This linear increase in $\alpha_1$ suggests either an increase in the number of events or an increased impact of these events. 

The different evolutions of these two fractional branches are typical of thermorheologically complex systems (or one might here say plethorheologically complex) due here to the distinct rheologies of liquid droplets in a solid \cite{plazek_1995_1996}. To simplify the analysis, we choose to first focus on the similar behaviors in the continuous-phase-dominated region at high frequency before comparing the variations due to the liquid droplets.

\subsection{Time-volume fraction superposition}
\begin{figure*}[hbt!]
	\hspace{-1cm}
    \includegraphics[width=15.75cm,keepaspectratio]{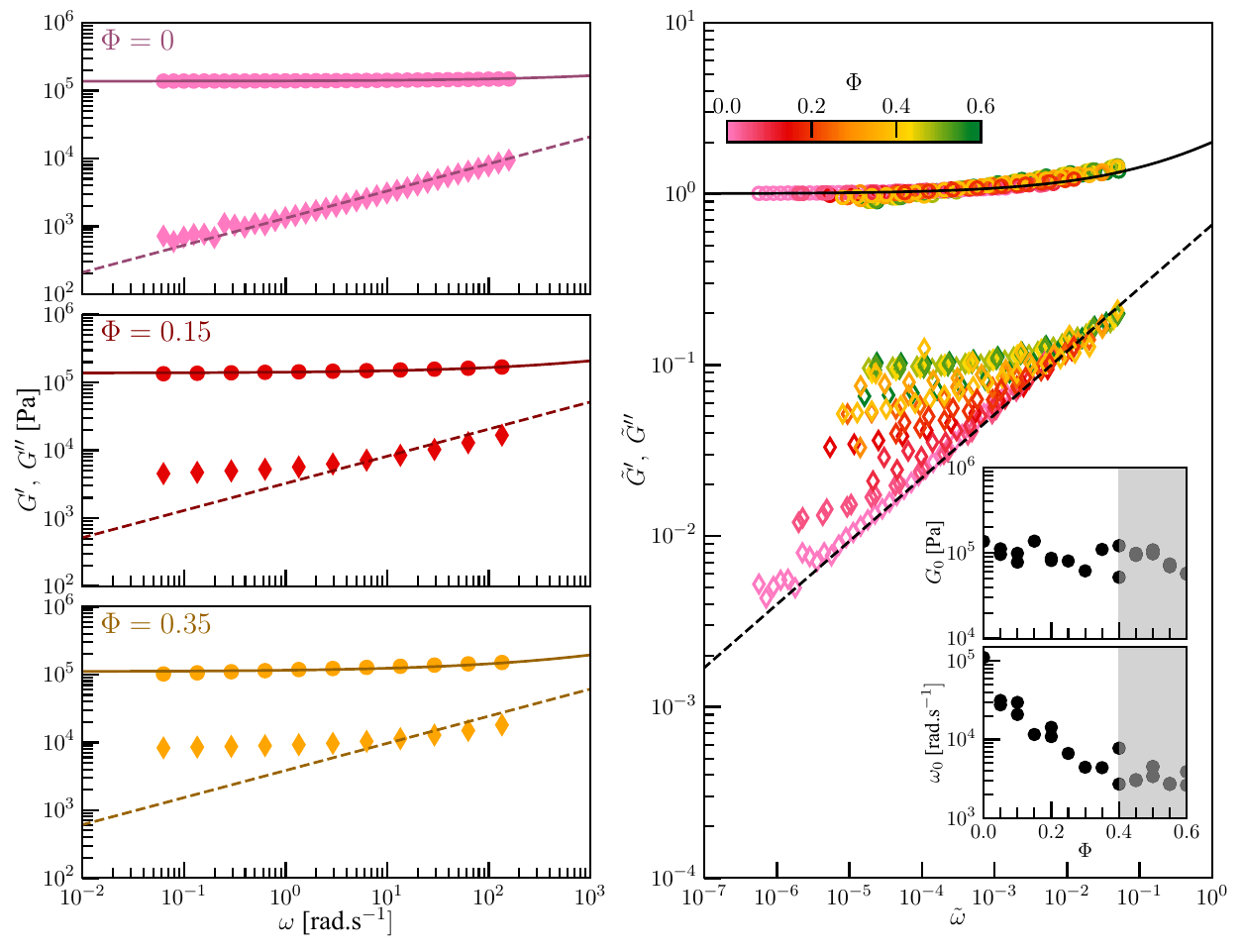}
	\caption{
	    \textbf{Left}  Storage ($\CIRCLE$) and loss ($\blacklozenge$) moduli evolution with  angular frequency for three samples of volume fraction $\Phi=0$ (top), $0.1$ (middle), and $0.25$ (bottom). The high frequency evolution has been fitted with a FKV0 model.
        \textbf{Right} Rescaled moduli for all samples. The superposition was obtained from the FKV0 fits, with $\tilde\Gp=\Gp/G_0$, $\tilde\Gpp=\Gpp/G_0$, and $\tilde\omega=\omega/\omega_0$ where $\omega_0=(G_0/(\mathds{V}_0\sin(2\alpha_0/\pi)))^{1/\alpha_0}$. The black curves represent the normalized FKV0 model $\tilde G^*=1+\mathds{V}_0/G_0(\ii\tilde\omega)^\alpha$, with $\mathds{V}_0/G_0=1.2$ and $\alpha=0.37$. $G_0$ and $\omega_0$ are represented in the inset as functions of the volume fraction. The grayed-out area points a change in behavior, especially visible on $\omega_0$, for $\Phi>0.4$.
		\label{fig:rescale}
		}
\end{figure*}

We fit all the samples with a FKV0 model (equation \ref{eq:FKV0}), to capture their frequency-evolution in the continuous-phase-dominated region. An acceptable fit is obtained for all high frequency moduli of the emulsions by setting $\alpha=0.4$. We can thus recover an elastic part due to the continuous phase $G_0$, and a characteristic frequency 
\begin{equation*}
    \omega_0 = \left(\frac{G_0}{\mathds{V}_0\sin(\tfrac{2}{\pi}\alpha_0)}\right)^{\frac{1}{\alpha_0}}
\end{equation*}
which we use to rescale the moduli of the samples such that $\tilde\Gp=\Gp/G_0$ and $\tilde\Gpp=\Gpp/G_0$ all follow the same rescaled fractional Kelvin-Voigt model of $\tilde\omega=\omega/\omega_0$. Some fits and the renormalized moduli are represented on figure \ref{fig:rescale}.

As more and more droplets are included in the samples, their influence on the rheology becomes more and more obvious, especially on the loss modulus. Our experiments show a stronger divergence from the continuous-phase-dominated region due to the droplets for $\Phi\ge0.15$ as is shown on figure \ref{fig:rescale} (Left). The high-frequency part of the spectra still superimpose well on the normalized general FKV0 model on figure \ref{fig:rescale} (right), which ranges over seven orders of magnitude, at least for volume fractions $\Phi\le0.4$. For emulsions of higher volume fraction however, the storage modulus seems to depart from the general model, coinciding with change in evolution of $\omega_0$ (insert). This change for higher volume fractions is coherent with the evolution observed for FKV1 on figure \ref{fig:FKV}. It might be due to either the experimental frequency interval being too small to properly observe the transition from droplet-dominated to continuous-phase-dominated regimes, or to a structure transition of the emulsion caused by the high volume fraction. Further investigation is required to interpret these points.

Interestingly, $G_0$ in our study is relatively constant around $10^5~\pascal$ while $\omega_0$ decreases with volume fraction.This is different from results in the literature on carbon black gels where a similar rescaling is used and both $G_0$ and $\omega_0$ were found to increase with volume fraction \cite{trappe_scaling_2000,legrand_dual_2023}. This is because the physical origin of the scaling is different. In our system, while the plateau modulus remains largely unaffected by the presence of droplets, the loss modulus clearly increases due to the dissipating liquid inclusions, leading to an earlier crossover. In the case of carbon black, the situation is opposite: the addition of particles leads to a percolating solid structure which extends the storage plateau modulus towards higher frequencies, and increases the stiffness of the material.

As the influence of the continuous phase on the relaxation spectrum has been normalized for most of the samples, we can now study the droplet-driven variation between samples. We focus on a given abscissa $\tilde\omega$ in the droplet-dominated region, before the transition to the continuous-phase-like behavior. We can compare the viscous dissipation in the samples while making sure that only the droplet-dominated behavior is at play. Whereas at a fixed angular frequency $\omega$, samples could be in different regimes, and thus the results could mix behavior in the droplet-dominated region, the transition region, and even from the continuous-phase-dominated region.

\subsection{Discussion of the master-curve}
To compare our samples at a given rescaled frequency $\tilde\omega$, we interpolate the experimental data with a "closest-fit" curve following FKV1 (equation \ref{eq:FKV1}). In order to stay coherent with the data, we will limit the study to the normalized frequencies that were accessible through the experiment, namely $\tilde\omega\in[10^{-7},\ 10^{-1}]$.

\begin{figure}[hbt!]
	\centering
	\includegraphics[width=8cm,keepaspectratio]{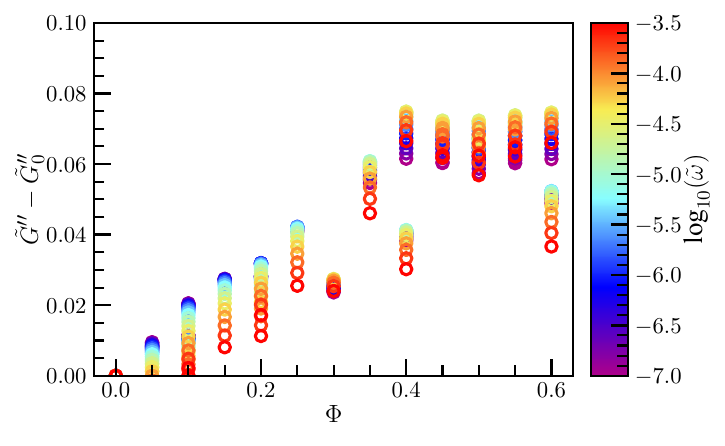}
	\caption{
	Evolution of the difference in dissipation between the emulsions ($\tilde\Gpp$) and the continuous phase ($\tilde\Gpp_0$) with volume fraction at different rescaled frequencies $\tilde\omega$ in the droplet-dominated frequency region.
		\label{fig:comparisons}
        }
\end{figure}

In order to study the influence of the liquid inclusions on the dissipative behavior of the solid emulsions, we focus on the droplet-dominated-regime. In this region, we subtract the evolution of the pure continuous phase from that of the composites to quantify the divergence from the global FKV0 model observed on figure \ref{fig:rescale}, such that we measure $\tilde\Gpp-\tilde\Gpp_0$ as functions of both the normalized frequency $\tilde\omega$ and the volume fraction. This evolution in the influence of the droplets is represented on figure \ref{fig:comparisons}.

As was already suspected from the decrease of $\omega_0$ with volume fraction, the dissipative behavior due to the droplets increases with $\Phi$, and saturates for $\Phi>0.4$. This increase is strikingly linear, and is independent of the rescaled frequency: the dissipative contributions from the continuous phase and the droplets thus seem to simply be additive. We can note that the saturation once again is coherent with the shift observed previously in FKV1 and in $\omega_0$. This additive evolution is reminiscent of the linearized equations for the volume-fraction-dependent rheology of dispersed systems $\eta_\mathrm{r} = 1 + k\Phi$ where $k$ is a constant with values typically between 1 and 2.5 \cite{einstein_motion_1905,krieger_mechanism_1959,faroughi_generalized_2015}. Thus, it would seem that at low frequencies, solid emulsions have an emulsion-like volume-fraction-dependent rheology, while at high frequencies they have a gel-like behavior tending towards the continuous phase.

\section{Conclusion}
We have studied the rheological response of soft solid emulsions formed of drops of PEG in PDMS, taking into account both the stiffness and the dissipative behavior. Our materials show a weak increase in storage modulus close to the elastomeric plateau of the pure continuous phase, and two asymptotic behaviors in loss modulus. At high frequency, the emulsions' loss modulus increases like the loss modulus of the continuous phase, whereas it is flattened at low frequencies, showing the influence of the encapsulated droplets. This behavior does not fit the usual model used to characterize viscoelastic blends, established by Palierne. We thus employed complementary analytical methods commonly used in the field of rheology of dispersed systems to better understand the response of the solid emulsions, and specifically their dissipative behavior.

A first fractional Kelvin-Voigt-based model shows the rheological complexity of the material, with two branches of independent evolution. We derive from it that the continuous phase dominates the rheology of the material at high frequencies, while the droplets dominate the lower frequencies due to their long relaxation time.

A second approach, inspired by the rheology of colloidal gels, allows to create a master-curve by rescaling the moduli based on the continuous-phase-dominated region. Due to the liquid inclusions, the loss moduli show a departure from this master-curve at low rescaled frequencies which can be compared to the dissipative behavior of the continuous phase. We find an evolution of the dissipative behavior proportional to the volume fraction of liquid in the samples, that then reaches a plateau for $\Phi>0.4$, independently of  the frequency. On the other hand, the storage modulus of the solid emulsions very closely resembles that of the continuous phase over the whole spectrum, for $\Phi\le0.4$.

These findings deepen our understanding of the intricate interactions in soft solid emulsion materials and provide essential guidance for developing new materials. Future investigations could examine the impact of various types of liquid inclusions and their interactions with the elastomer matrix, further advancing our knowledge and optimizing material performance. Future works shall include creep and relaxation studies to confirm the validity of the fractional models, as well as simulations to better understand the cause of the seemingly independent variations of the elastic and dissipative responses.

Furthermore, it would be interesting to establish a more complete theoretical framework to study this class of systems, whose rheology seems to bridge the gap between emulsions and gels.

\section*{Author contributions}
\noindent Conceptualization: E.G., A.S. and C.P.;
Data curation: E.G.;
Formal analysis: E.G. and C.P.;
Funding Acquisition: C.P.;
Investigation: E.G.;
Methodology: E.G. and C.P.;
Project administration: A.S. and C.P.;
Resource: C.P.;
Software: E.G.;
Supervision: A.S. and C.P.;
Validation: E.G. and C.P.;
Visualization: E.G.;
Writing -- original draft: E.G.;
Writing -- review \& editing: E.G., A.S. and C.P.

\section*{Conflict of interest}
\noindent
The authors declare no competing financial interest.

\section*{Acknowledgments}
\noindent
The authors thank Sandrine Mariot and Laura Wallon for their help in the experiments and the sample generation.
We would also like to acknowledge Emanuela Del Gado for the fascinating discussion on building master curves using fractional rheology to reveal the influence of the structure of soft materials. We are also very thankful to Thibaut Divoux for the enlightening discussion on fractional fitting and rescaling we shared at the AERC.

\section*{Data availability}
\noindent
The data for this study are available at the following DOI : \href{https://doi.org/10.5281/zenodo.17131322}{10.5281/zenodo.17131322}.

\bibliographystyle{elsarticle-num} 
\bibliography{references.bib}

\begin{thebibliography}{10}
\expandafter\ifx\csname url\endcsname\relax
  \def\url#1{\texttt{#1}}\fi
\expandafter\ifx\csname urlprefix\endcsname\relax\def\urlprefix{URL }\fi
\expandafter\ifx\csname href\endcsname\relax
  \def\href#1#2{#2} \def\path#1{#1}\fi

\bibitem{nutting_new_1921}
P.~G. Nutting, A new general law of deformation, Journal of the Franklin
  Institute 191~(5) (1921) 679--685.
\newblock \href {https://doi.org/10.1016/s0016-0032(21)90171-6}
  {\path{doi:10.1016/s0016-0032(21)90171-6}}.

\bibitem{michelson_laws_1917}
A.~Michelson, The {{Laws}} of {{Elastico-Viscous Flow}}, Proceedings of the
  National Academy of Sciences 3~(5) (1917).

\bibitem{williams_temperature_1955}
M.~L. Williams, R.~F. Landel, J.~D. Ferry, The temperature dependence of
  relaxation mechanisms in amorphous polymers and other glass-forming liquids,
  Journal of the American Chemical Society 77~(14) (1955) 3701--3707.
\newblock \href {https://doi.org/10.1021/ja01619a008}
  {\path{doi:10.1021/ja01619a008}}.

\bibitem{blair_subjective_1939}
G.~W.~S. Blair, F.~M.~V. Coppen, The subjective judgement of the elastic and
  plastic properties of soft bodies; the "differential thresholds" for
  viscosities and compression moduli, Proceedings of the Royal Society of
  London. Series B - Biological Sciences 128~(850) (1939) 109--125.
\newblock \href {https://doi.org/10.1098/rspb.1939.0046}
  {\path{doi:10.1098/rspb.1939.0046}}.

\bibitem{blair_analytical_1944}
G.~W.~S. Blair, Analytical and integrative aspects of the stress-strain-time
  problem, Journal of Scientific Instruments 21~(5) (1944) 80--84.
\newblock \href {https://doi.org/10.1088/0950-7671/21/5/302}
  {\path{doi:10.1088/0950-7671/21/5/302}}.

\bibitem{blair_limitations_1947}
G.~W.~S. Blair, B.~C. Veinoglou, Limitations of the {{Newtonian}} time scale in
  relation to non-equilibrium rheological states and a theory of
  quasi-properties, Proceedings of the Royal Society of London. Series A.
  Mathematical and Physical Sciences 189~(1016) (1947) 69--87.
\newblock \href {https://doi.org/10.1098/rspa.1947.0029}
  {\path{doi:10.1098/rspa.1947.0029}}.

\bibitem{caputo_linear_1967}
M.~Caputo, Linear models of dissipation whose q is almost frequency
  independent--ii, Geophysical Journal International 13~(5) (1967) 529--539.
\newblock \href {https://doi.org/10.1111/j.1365-246x.1967.tb02303.x}
  {\path{doi:10.1111/j.1365-246x.1967.tb02303.x}}.

\bibitem{bagley_theoretical_1983}
R.~L. Bagley, P.~J. Torvik, A theoretical basis for the application of
  fractional calculus to viscoelasticity, Journal of Rheology 27~(3) (1983)
  201--210.
\newblock \href {https://doi.org/10.1122/1.549724}
  {\path{doi:10.1122/1.549724}}.

\bibitem{bagley_fractional_1986}
R.~L. Bagley, P.~J. Torvik, On the fractional calculus model of viscoelastic
  behavior, Journal of Rheology 30~(1) (1986) 133--155.
\newblock \href {https://doi.org/10.1122/1.549887}
  {\path{doi:10.1122/1.549887}}.

\bibitem{nonnenmacher_fractional_1991}
T.~F. Nonnenmacher, W.~G. Glöckle, A fractional model for mechanical stress
  relaxation, Philosophical Magazine Letters 64~(2) (1991) 89--93.
\newblock \href {https://doi.org/10.1080/09500839108214672}
  {\path{doi:10.1080/09500839108214672}}.

\bibitem{gloeckle_fractional_1991}
W.~G. Gloeckle, T.~F. Nonnenmacher, Fractional integral operators and fox
  functions in the theory of viscoelasticity, Macromolecules 24~(24) (1991)
  6426--6434.
\newblock \href {https://doi.org/10.1021/ma00024a009}
  {\path{doi:10.1021/ma00024a009}}.

\bibitem{gloeckle_fractional_1994}
W.~G. Gloeckle, T.~F. Nonnenmacher, Fractional relaxation and the
  time-temperature superposition principle, Rheologica Acta 33~(4) (1994)
  337--343.
\newblock \href {https://doi.org/10.1007/bf00366960}
  {\path{doi:10.1007/bf00366960}}.

\bibitem{schiessel_hierarchical_1993}
H.~Schiessel, A.~Blumen, Hierarchical analogues to fractional relaxation
  equations, Journal of Physics A: Mathematical and General 26~(19) (1993)
  5057--5069.
\newblock \href {https://doi.org/10.1088/0305-4470/26/19/034}
  {\path{doi:10.1088/0305-4470/26/19/034}}.

\bibitem{schiessel_generalized_1995}
H.~Schiessel, R.~Metzler, A.~Blumen, T.~F. Nonnenmacher, Generalized
  viscoelastic models: their fractional equations with solutions, Journal of
  Physics A: Mathematical and General 28~(23) (1995) 6567--6584.
\newblock \href {https://doi.org/10.1088/0305-4470/28/23/012}
  {\path{doi:10.1088/0305-4470/28/23/012}}.

\bibitem{jaishankar_fractional_2014}
A.~Jaishankar, G.~H. McKinley, A fractional k-bkz constitutive formulation for
  describing the nonlinear rheology of multiscale complex fluids, Journal of
  Rheology 58~(6) (2014) 1751--1788.
\newblock \href {https://doi.org/10.1122/1.4892114}
  {\path{doi:10.1122/1.4892114}}.

\bibitem{jaishankar_power_2013}
A.~Jaishankar, G.~H. McKinley, Power-law rheology in the bulk and at the
  interface: quasi-properties and fractional constitutive equations,
  Proceedings of the Royal Society A: Mathematical, Physical and Engineering
  Sciences 469~(2149) (2013) 20120284.
\newblock \href {https://doi.org/10.1098/rspa.2012.0284}
  {\path{doi:10.1098/rspa.2012.0284}}.

\bibitem{bantawa_hidden_2023}
M.~Bantawa, B.~Keshavarz, M.~Geri, M.~Bouzid, T.~Divoux, G.~H. McKinley,
  E.~Del~Gado, The hidden hierarchical nature of soft particulate gels, Nature
  Physics 19~(8) (2023) 1178--1184.
\newblock \href {https://doi.org/10.1038/s41567-023-01988-7}
  {\path{doi:10.1038/s41567-023-01988-7}}.

\bibitem{faber_describing_2017}
T.~J. Faber, A.~Jaishankar, G.~H. McKinley, Describing the firmness,
  springiness and rubberiness of food gels using fractional calculus. part ii:
  Measurements on semi-hard cheese, Food Hydrocolloids 62 (2017) 325--339.
\newblock \href {https://doi.org/10.1016/j.foodhyd.2016.06.038}
  {\path{doi:10.1016/j.foodhyd.2016.06.038}}.

\bibitem{faber_describing_2017a}
T.~J. Faber, A.~Jaishankar, G.~H. McKinley, Describing the firmness,
  springiness and rubberiness of food gels using fractional calculus. part i:
  Theoretical framework, Food Hydrocolloids 62 (2017) 311--324.
\newblock \href {https://doi.org/10.1016/j.foodhyd.2016.05.041}
  {\path{doi:10.1016/j.foodhyd.2016.05.041}}.

\bibitem{faber_firm_2017}
T.~J. Faber, L.~C.~A. Van~Breemen, G.~H. McKinley, From firm to fluid –
  structure-texture relations of filled gels probed under large amplitude
  oscillatory shear, Journal of Food Engineering 210 (2017) 1--18.
\newblock \href {https://doi.org/10.1016/j.jfoodeng.2017.03.028}
  {\path{doi:10.1016/j.jfoodeng.2017.03.028}}.

\bibitem{style_solid–liquid_2020}
R.~W. Style, R.~Tutika, J.~Y. Kim, M.~D. Bartlett, Solid–liquid composites
  for soft multifunctional materials, Advanced Functional Materials 31~(1)
  (2020) 2005804.
\newblock \href {https://doi.org/10.1002/adfm.202005804}
  {\path{doi:10.1002/adfm.202005804}}.

\bibitem{sheng_solid–liquid_2021}
Z.~Sheng, Y.~Ding, G.~Li, C.~Fu, Y.~Hou, J.~Lyu, K.~Zhang, X.~Zhang,
  Solid–liquid host–guest composites: The marriage of porous solids and
  functional liquids, Advanced Materials 33~(52) (2021) 2104851.
\newblock \href {https://doi.org/10.1002/adma.202104851}
  {\path{doi:10.1002/adma.202104851}}.

\bibitem{yu_solid–liquid_2021}
M.~Yu, X.~Li, P.~Lv, H.~Duan, Solid–liquid composites with high impact
  resistance, Acta Mechanica Solida Sinica 34~(6) (2021) 911--921.
\newblock \href {https://doi.org/10.1007/s10338-021-00277-1}
  {\path{doi:10.1007/s10338-021-00277-1}}.

\bibitem{li_solid–liquid_2023}
X.~Li, Y.~Liu, Y.~Xu, P.~Tao, T.~Deng, Solid–liquid phase change composite
  materials for direct solar–thermal energy harvesting and storage, Accounts
  of Materials Research 4~(6) (2023) 484--495.
\newblock \href {https://doi.org/10.1021/accountsmr.2c00251}
  {\path{doi:10.1021/accountsmr.2c00251}}.

\bibitem{cai_printing_2020}
L.~Cai, J.~Marthelot, C.~Falcón, P.~M. Reis, P.-T. Brun, Printing on liquid
  elastomers, Soft Matter 16~(12) (2020) 3137--3142.
\newblock \href {https://doi.org/10.1039/c9sm02452b}
  {\path{doi:10.1039/c9sm02452b}}.

\bibitem{langendorfer_viscoelastic_2024}
L.~J. Langendörfer, E.~Guseva, P.~Bauermann, A.~Schubert, O.~Hensel,
  M.~Diakité, \href{https://www.mdpi.com/2304-8158/13/23/3875}{The
  viscoelastic behavior of legume protein emulsion gels—the effect of heating
  temperature and oil content on viscoelasticity, the degree of networking, and
  the microstructure}, Foods 13~(23) (2024).
\newblock \href {https://doi.org/10.3390/foods13233875}
  {\path{doi:10.3390/foods13233875}}.
\newline\urlprefix\url{https://www.mdpi.com/2304-8158/13/23/3875}

\bibitem{ford_multifunctional_2019}
M.~J. Ford, C.~P. Ambulo, T.~A. Kent, E.~J. Markvicka, C.~Pan, J.~Malen, T.~H.
  Ware, C.~Majidi, A multifunctional shape-morphing elastomer with liquid metal
  inclusions, Proceedings of the National Academy of Sciences 116~(43) (2019)
  21438--21444.

\bibitem{ma_soft_2017}
Y.~Ma, M.~Pharr, L.~Wang, J.~Kim, Y.~Liu, Y.~Xue, R.~Ning, X.~Wang, H.~U.
  Chung, X.~Feng, et~al., Soft elastomers with ionic liquid-filled cavities as
  strain isolating substrates for wearable electronics, small 13~(9) (2017)
  1602954.

\bibitem{yun_liquid_2019}
G.~Yun, S.-Y. Tang, S.~Sun, D.~Yuan, Q.~Zhao, L.~Deng, S.~Yan, H.~Du, M.~D.
  Dickey, W.~Li, Liquid metal-filled magnetorheological elastomer with positive
  piezoconductivity, Nature communications 10~(1) (2019) 1300.

\bibitem{ankit_high_2020}
Ankit, N.~Tiwari, F.~Ho, F.~Krisnadi, M.~R. Kulkarni, L.~L. Nguyen, S.~J.~A.
  Koh, N.~Mathews, \href{https://doi.org/10.1021/acsami.0c08754}{High-k,
  ultrastretchable self-enclosed ionic liquid-elastomer composites for soft
  robotics and flexible electronics}, ACS Applied Materials \& Interfaces
  12~(33) (2020) 37561--37570, pMID: 32814378.
\newblock \href {http://arxiv.org/abs/https://doi.org/10.1021/acsami.0c08754}
  {\path{arXiv:https://doi.org/10.1021/acsami.0c08754}}, \href
  {https://doi.org/10.1021/acsami.0c08754} {\path{doi:10.1021/acsami.0c08754}}.
\newline\urlprefix\url{https://doi.org/10.1021/acsami.0c08754}

\bibitem{peng_highly_2023}
M.~Peng, B.~Ma, G.~Li, Y.~Liu, Y.~Zhang, X.~Ma, S.~Yan, A highly stretchable
  and sintering-free liquid metal composite conductor enabled by ferrofluid,
  Soft Science 3~(4) (2023) 1--12.

\bibitem{style_stiffening_2014}
R.~W. Style, R.~Boltyanskiy, B.~Allen, K.~E. Jensen, H.~P. Foote,
  J.~Wettlaufer, E.~R. Dufresne, Stiffening solids with liquid inclusions,
  Nature Physics 11~(1) (2014) 82--87.
\newblock \href {https://doi.org/10.1038/nphys3181}
  {\path{doi:10.1038/nphys3181}}.

\bibitem{ghosh_elastomers_2022}
K.~Ghosh, O.~Lopez-Pamies, Elastomers filled with liquid inclusions: Theory,
  numerical implementation, and some basic results, Journal of the Mechanics
  and Physics of Solids 166 (2022) 104930.
\newblock \href {https://doi.org/10.1016/j.jmps.2022.104930}
  {\path{doi:10.1016/j.jmps.2022.104930}}.

\bibitem{ghasemi_multiscale_2021}
H.~Ghasemi, A multiscale material model for heterogeneous liquid droplets in
  solid soft composites, Frontiers of Structural and Civil Engineering 15~(5)
  (2021) 1292--1299.
\newblock \href {https://doi.org/10.1007/s11709-021-0771-3}
  {\path{doi:10.1007/s11709-021-0771-3}}.

\bibitem{hamdia_quantifying_2022}
K.~M. Hamdia, H.~Ghasemi, Quantifying the uncertainties in modeling soft
  composites via a multiscale approach, International Journal of Solids and
  Structures 256 (2022) 111959.
\newblock \href {https://doi.org/10.1016/j.ijsolstr.2022.111959}
  {\path{doi:10.1016/j.ijsolstr.2022.111959}}.

\bibitem{gilbert_decoupling_2024}
E.~Gilbert, A.~Salonen, C.~Poulard, Decoupling the rheological responses of a
  soft solid emulsion with liquid inclusions, Journal of Physics: Condensed
  Matter 36~(42) (2024) 425103.
\newblock \href {https://doi.org/10.1088/1361-648x/ad61ad}
  {\path{doi:10.1088/1361-648x/ad61ad}}.

\bibitem{palierne_linear_1990}
J.~F. Palierne, Linear rheology of viscoelastic emulsions with interfacial
  tension, Rheologica Acta 29~(3) (1990) 204--214.
\newblock \href {https://doi.org/10.1007/bf01331356}
  {\path{doi:10.1007/bf01331356}}.

\bibitem{pal_new_2008}
R.~Pal, A new linear viscoelastic model for emulsions and suspensions, Polymer
  Engineering \& Science 48~(7) (2008) 1250--1253.
\newblock \href {https://doi.org/10.1002/pen.21065}
  {\path{doi:10.1002/pen.21065}}.

\bibitem{curro_theoretical_1983}
J.~G. Curro, P.~Pincus, A theoretical basis for viscoelastic relaxation of
  elastomers in the long-time limit, Macromolecules 16~(4) (1983) 559--562.

\bibitem{sollich_rheology_1997}
P.~Sollich, F.~Lequeux, P.~Hébraud, M.~E. Cates, Rheology of soft glassy
  materials, Physical Review Letters 78~(10) (1997) 2020--2023.
\newblock \href {https://doi.org/10.1103/physrevlett.78.2020}
  {\path{doi:10.1103/physrevlett.78.2020}}.

\bibitem{sollich_rheological_1998}
P.~Sollich, Rheological constitutive equation for a model of soft glassy
  materials, Physical Review E 58~(1) (1998) 738--759.
\newblock \href {https://doi.org/10.1103/physreve.58.738}
  {\path{doi:10.1103/physreve.58.738}}.

\bibitem{hebraud_role_2000}
P.~Hébraud, F.~Lequeux, J.-F. Palierne, Role of permeation in the linear
  viscoelastic response of concentrated emulsions, Langmuir 16~(22) (2000)
  8296--8299.
\newblock \href {https://doi.org/10.1021/la001091g}
  {\path{doi:10.1021/la001091g}}.

\bibitem{creton_fracture_2016}
C.~Creton, M.~Ciccotti, Fracture and adhesion of soft materials: a review,
  Reports on Progress in Physics 79~(4) (2016) 046601.
\newblock \href {https://doi.org/10.1088/0034-4885/79/4/046601}
  {\path{doi:10.1088/0034-4885/79/4/046601}}.

\bibitem{plazek_1995_1996}
D.~J. Plazek, 1995 bingham medal address: Oh, thermorheological simplicity,
  wherefore art thou?, Journal of Rheology 40~(6) (1996) 987--1014.
\newblock \href {https://doi.org/10.1122/1.550776}
  {\path{doi:10.1122/1.550776}}.

\bibitem{trappe_scaling_2000}
V.~Trappe, D.~A. Weitz, Scaling of the viscoelasticity of weakly attractive
  particles, Physical Review Letters 85~(2) (2000) 449--452.
\newblock \href {https://doi.org/10.1103/physrevlett.85.449}
  {\path{doi:10.1103/physrevlett.85.449}}.

\bibitem{legrand_dual_2023}
G.~Legrand, S.~Manneville, G.~H. McKinley, T.~Divoux, Dual origin of
  viscoelasticity in polymer-carbon black hydrogels: A rheometry and electrical
  spectroscopy study, Macromolecules 56~(6) (2023) 2298--2308.
\newblock \href {https://doi.org/10.1021/acs.macromol.2c02068}
  {\path{doi:10.1021/acs.macromol.2c02068}}.

\bibitem{einstein_motion_1905}
A.~Einstein, On the {{Motion}} of {{Small Particles Suspended}} in {{Liquids}}
  at {{Rest Required}} by the {{Molecular-Kinetic Theory}} of {{Heat}}, in:
  Investigation of the {{Theory}} of {{Brownian}} Movement, dover publications
  Edition, 1905, p. 137.

\bibitem{krieger_mechanism_1959}
I.~M. Krieger, T.~J. Dougherty, A {{Mechanism}} for {{Non}}-{{Newtonian Flow}}
  in {{Suspensions}} of {{Rigid Spheres}}, Transactions of the Society of
  Rheology 3~(1) (1959) 137--152.
\newblock \href {https://doi.org/10.1122/1.548848}
  {\path{doi:10.1122/1.548848}}.

\bibitem{faroughi_generalized_2015}
S.~A. Faroughi, C.~Huber, A generalized equation for rheology of emulsions and
  suspensions of deformable particles subjected to simple shear at low
  {{Reynolds}} number, Rheol Acta 54~(2) (2015) 85--108.
\newblock \href {https://doi.org/10.1007/s00397-014-0825-8}
  {\path{doi:10.1007/s00397-014-0825-8}}.

\end{thebibliography}

\end{document}